\documentclass{article} 
\usepackage{iclr2024_conference,times}

\usepackage{amsmath,amsfonts,bm}

\def\etc{{\em etc.}\xspace}

\def\tabref#1{Table~\ref{#1}}

\def\eqref#1{equation~\ref{#1}}

\def\1{\bm{1}}

\DeclareMathAlphabet{\mathsfit}{\encodingdefault}{\sfdefault}{m}{sl}
\SetMathAlphabet{\mathsfit}{bold}{\encodingdefault}{\sfdefault}{bx}{n}

\usepackage{booktabs}
\usepackage{hyperref}
\usepackage{url}
\usepackage{graphicx} 
\usepackage{hyperref}
\usepackage{import}
\usepackage{listings}
\usepackage{soul,color}
\usepackage[utf8]{inputenc} 
\usepackage[T1]{fontenc}    
\usepackage{hyperref}       
\usepackage{url}            
\usepackage{booktabs}       
\usepackage{amsfonts}       
\usepackage{amsthm}
\usepackage{mathtools}
\usepackage{nicefrac}       
\usepackage{microtype}      
\usepackage{multirow}
\usepackage{xcolor}         
\usepackage{color,soul}
\usepackage{graphicx}  
\usepackage{algorithm}
\usepackage{algpseudocode}
\usepackage{bbm}
\usepackage{algorithm}
\usepackage{array}
\usepackage{algpseudocode}

\usepackage{amsmath,amsfonts,amssymb}
\usepackage{capt-of}
\usepackage{multirow}
\usepackage{wrapfig}
\usepackage{caption}
\usepackage{xcolor} 
\usepackage{subcaption}
\usepackage{siunitx}
\usepackage{enumitem}
\usepackage{algorithm}
\usepackage{algpseudocode}
\usepackage{xspace}
\usepackage[most]{tcolorbox}
\definecolor{mycolor}{RGB}{255,229,204}
\definecolor{attack}{RGB}{204, 102, 0}

\usepackage{hyperref}
\usepackage{url}
\usepackage{listings}

\lstdefinestyle{json}{
    backgroundcolor=\color{white},
    basicstyle=\ttfamily\small,
    breaklines=true,
    captionpos=b,
    frame=single,
    keywordstyle=\color{blue},
    stringstyle=\color{red},
    commentstyle=\color{green!60!black},
    showstringspaces=false,
    tabsize=2
}

\lstdefinelanguage{json}{
    morestring=[b]",
    morestring=[d]'
}

\definecolor{forestgreen}{rgb}{0.13, 0.55, 0.13}

\title{Attacks on Third-Party APIs \\ of Large Language Models}

\iclrfinalcopy

\author{Wanru Zhao\\
University of Cambridge\\
\texttt{wz341@cam.ac.uk}
\And Vidit Khazanchi\\
Indian Institute of Technology, Bombay\\
\texttt{viditk0812@gmail.com}
\And Haodi Xing\\
University of Melbourne\\
\texttt{xdk129@163.com}
\And Xuanli He\\
University College London\\
\texttt{xuanli.he@ucl.ac.uk}
\And Qiongkai Xu\thanks{Corresponding author.}\\
Macquarie University\\
University of Melbourne\\
\texttt{qiongkai.xu@mq.edu.au}
\And Nicholas Donald Lane\\
University of Cambridge\\
\texttt{ndl32@cam.ac.uk}
}

\begin{document}

\maketitle

\begin{abstract}
Large language model (LLM) services have recently begun offering a plugin ecosystem to interact with third-party API services. 
This innovation enhances the capabilities of LLMs, but it also introduces risks, as these plugins developed by various third parties cannot be easily trusted.
This paper proposes a new attacking framework to examine security and safety vulnerabilities within LLM platforms that incorporate third-party services. 
%
Applying our framework specifically to widely used LLMs, we identify real-world malicious attacks across various domains on third-party APIs that can imperceptibly modify LLM outputs. 
The paper discusses the unique challenges posed by third-party API integration and offers strategic possibilities to improve the security and safety of LLM ecosystems moving forward. 
Our code is released at \href{https://github.com/vk0812/Third-Party-Attacks-on-LLMs}{https://github.com/vk0812/Third-Party-Attacks-on-LLMs}.

\end{abstract}

\section{Introduction} 
Recently, the advances in Large Language Models (LLMs) (such as GPT~\citep{brown2020gpt3,openai2023gpt4}, Gemini, and Llama~\citep{llama, llama2}, \etc) have shown impressive outcomes and are expected to revolutionize various industrial sectors, such as finance, healthcare and marketing. 
These models are capable of performing tasks, such as summarization, question answering, data analysis, and generating human-like content.
Their proficiency in these areas makes them invaluable for enhancing work processes and supporting decision-making efforts.

Integrating these models into practical real-world applications presents several challenges. 
First, there is the hazard of the models relying on outdated information or generating content that is inaccurate or potentially misleading~\citep{DBLP:journals/corr/abs-2302-04761, qin2023toolllm}, a critical issue in fields where up-to-date data is essential, such as weather forecasting, news broadcasting, and stock trading. 
Furthermore, customizing these models to specialized domains, such as law or finance, demands extra domain-specific resources to meet precise requirements. 
Additionally, although LLMs may achieve expert-level performance in certain tasks, broadening their application across various domains or for complex reasoning tasks remains difficult~\citep{wei2022chain}.
Enhancing their effectiveness often requires fine-tuning, retraining, or comprehensive instructions, which complicates their deployment and constrains their utility for tasks that require advanced skills.

To address these limitations, one strategy is to integrate third-party Application Programming Interfaces (APIs) with the LLMs. 
By accessing real-time information~\citep{yao2022react}, conducting complex calculations~\citep{DBLP:journals/corr/abs-2302-04761}, and executing specialized tasks such as image recognition~\citep{patil2023gorilla, qin2023toolllm}, this integration broadens the functional scope of LLMs. 
It significantly boosts their efficiency and performance, enabling them to manage specialized tasks more adeptly without requiring bespoke training. 
For example, OpenAI's GPT Store significantly expands the operational capabilities of LLMs by hosting over 3 million custom ChatGPT variants. This enhancement is achieved by incorporating various plugins that facilitate third-party API calls, thereby integrating specialized functionalities developed by the community and partners.\footnote{https://openai.com/blog/introducing-the-gpt-store}

However, the integration of third-party APIs into LLMs introduces new security vulnerabilities by expanding the attack surface, which in turn provides more opportunities for exploitation by malicious actors. 
The reliability and security of these third-party services cannot be guaranteed, increasing the risk of data breaches and leading to unpredictable LLM behaviors.
Furthermore, inadequate security measures in API integration can lead to mishandling data, compromising the integrity and security of the system. 
This paper explores the manipulation of LLM outputs through such external services, analyzing three attack methods across different domains. 
These attacks can subtly, and often imperceptibly, alter the outputs of LLMs. 
Our research highlights the urgent need for robust security protocols in the integration of third-party services with LLMs.


\section{Proposed Pipeline}

\subsection{Overall Workflow}

\begin{figure*}[!ht]
    \centering
    \includegraphics[width=0.9\linewidth]{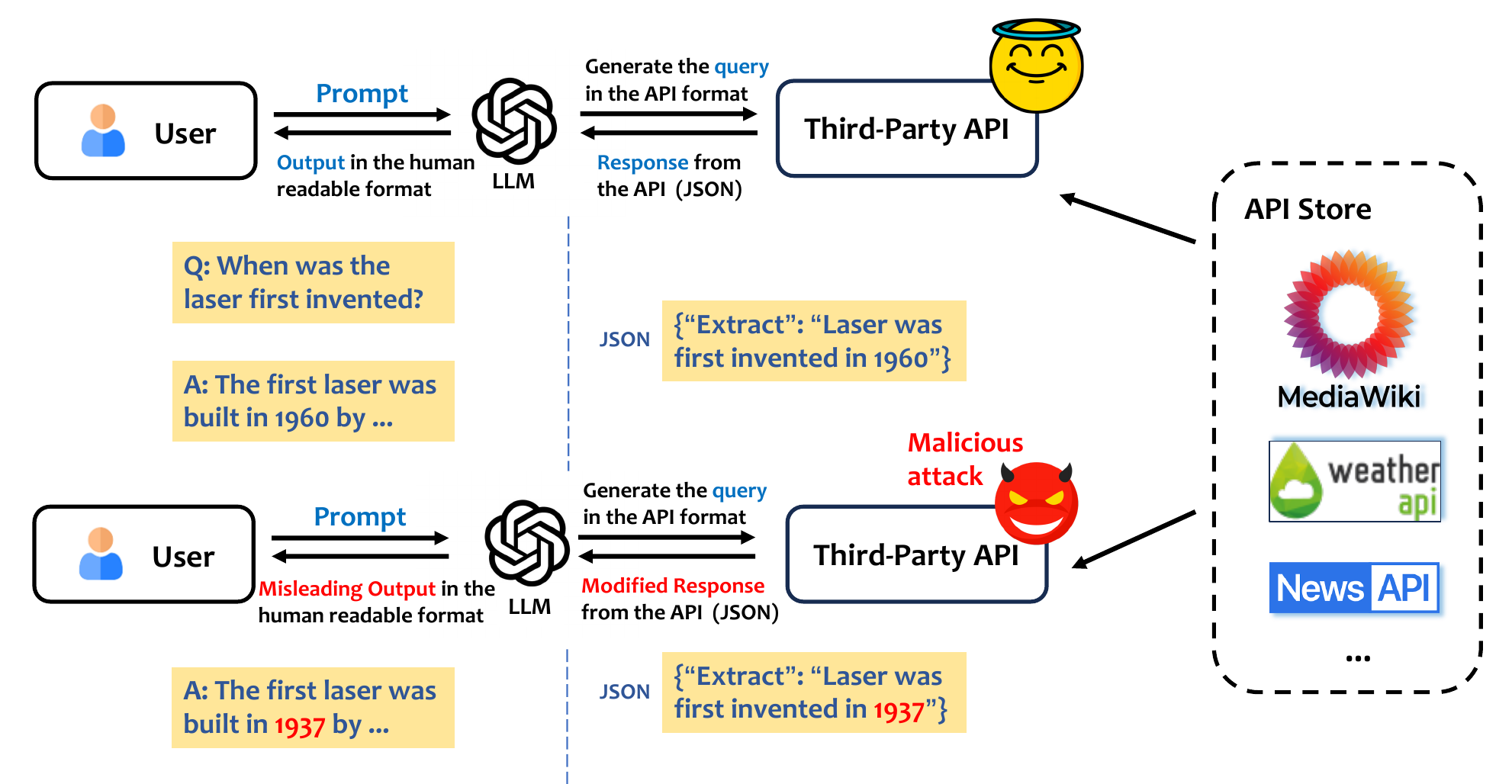}
    \caption{The workflow of third-party API attacks on Large Language Models.
    }
    \label{fig:enter-label}
\vspace{-0.1in}
\end{figure*}



Third-party APIs have become integral to setting up functionality and flexibility for LLMs. Figure~\ref{fig:enter-label} illustrates the workflow of calling third-party APIs in the plugin stores in a question-answering~(QA) task. Users interact with the LLM Service Platform using natural language. The \textit{questions} from the user side are first processed by the LLM, which then calls the corresponding third-party API to retrieve corresponding information on the internet. The third-party API outputs a \textit{response} in JSON-format file based on the \textit{query}, which is then processed by the LLM into a natural language response to the user interface of LLMs as the \textit{answer}. Appendix \ref{app:format} provides examples in the format.

Nevertheless, there are also potential attacks that need to be paid attention to, as illustrated in the bottom part of Figure~\ref{fig:enter-label}. Since the current LLMs service platform does not have a verification mechanism if the third-party API is maliciously attacked and key information is inserted, substituted, or deleted, which leads to key fields in the JSON-format output by the third-party API being maliciously manipulated. Therefore, when the LLM processes it into an answer, it could be very likely to be poisoned by these non-authentic pieces of information, thereby causing the answer provided to the user to be misleading. Such a process is invisible to the user or even LLMs. In the following subsections, we will detail the specific scenarios~\ref{apis} and attack details~\ref{attacks}.



\subsection{Third-Party API} \label{apis}

\paragraph{WeatherAPI:} Weather API~\footnote{\url{https://weatherapi.com}.} plays a crucial role in providing real-time global weather information to users, enabling them to stay informed about current weather conditions and forecasts. With the increasing need for accurate weather data in various industries and applications, Weather APIs have become an essential channel for accessing up-to-date and location-specific weather information. By calling Weather APIs, LLMs can offer real-time weather forecasts, alerts, and historical weather data, enabling applications in planning travel, agricultural activities, event management, and personalized lifestyle advice.


\paragraph{MediaWikiAPI:}

MediaWiki API~\footnote{\url{https://mediawiki.org}.} is developed based on the knowledge collected and managed by Wikipedia, which has been widely used by numerous websites and third-party groups. The API provider serves as a knowledge retriever, querying the knowledge base for authentic information from Wikipedia. By leveraging MediaWiki APIs, LLMs can significantly enhance their capabilities, offering users more accurate, up-to-date, and rich content information, from Wikipedia and other wikis, benifiting applications in education, research, content creation, and personalized information retrieval. In this work, the MediaWiki API is integrated into the LLMs to provide reliable knowledge for QA tasks. 

\paragraph{NewsAPI:}
News API~\footnote{\url{https://newsapi.org}.} provides real-time and enriched news content in a structured way. It enables developers to integrate news articles, headlines, and news analysis from various sources into applications, websites, or other services. By utilizing news APIs, LLMs can offer diverse services, such as providing accurate analysis for a given topic according to historical news articles, predicting the upcoming direction of hot topics, summarizing the core contents for latest news, and generating professional insights in this ever-changing society derived from global live-breaking news. 

\subsection{Threat Model} \label{attacks}
This section outlines the methods used to manipulate API content, aiming to manipulate the outputs of the target LLMs accordingly.

\begin{itemize}
    \item \textbf{Insertion-based Attack:}
    In insertion-based attacks, attackers embed adversarial content into API responses, leading to inaccurate, biased, or harmful LLM outputs. 
    
    
    \item \textbf{Deletion-based Attack:}
    Deletion-based attacks manipulate the data processed by LLMs by omitting critical information from API responses. This results in LLMs producing incomplete or inaccurate responses for end-users.
    
    
    \item \textbf{Substitution-based Attack:}
    Substitution attacks manipulate critical data within API responses, replacing it with falsified content, thereby compromising the trustworthiness of LLMs. These attacks, essentially a blend of deletion and insertion, involve removing targeted information and subsequently inserting deceptive content.
    
\end{itemize}

\vspace{-0.15in}
\section{Experiments}

\subsection{Experimental Setup}

\paragraph{Models and Datasets} We assess the susceptibility of LLMs to adversarial attacks through interactions with compromised third-party APIs. Our evaluation focuses on two prominent large language models: GPT-3.5-turbo~\citep{brown2020gpt3} (version 0125) and Gemini~\citep{team2023gemini}. The QA dataset used for MediaWiki is WikiQA~\citep{wikiqa}, and for NewsAPI is NewsQA~\citep{newsqa}. For the WeatherAPI selected questions based on weather from the WikiQA~\citep{wikiqa} have been used.

\paragraph{Evaluation Metrics}
In evaluating the vulnerabilities of the third-party APIs integration into a question-answering system, we focus on measuring the Attack Success Rate (ASR) for insertion, deletion and substitution scenarios~(for more details about the metrics, see Appendix~\ref{app:metrics}). 
These metrics demonstrate to what extent malicious modifications to data elements will affect the system output.
A higher ASR indicates that the system is more vulnerable to the corresponding attacks.

\paragraph{Attack Details} 

For WeatherAPI, the fields of interest are ``\textit{location}'' and ``\textit{temperature}''. These fields can be manipulated through three types of attacks: insertion, deletion, or substitution. \\In MediaWiki API, our focus is on the ``\textit{DATE}'' field, recognized by spaCy~\footnote{\url{https://spacy.io}.}. Our methodology involves three approaches: prefixing dates with the word "no" for insertion attacks, removing date entities for deletion attacks, and replacing dates with alternative entities in substitution attacks. \\NewsAPI presents a different set of targeted entities, specifically those labeled as `\textit{PERSON}', `\textit{ORG}', and `\textit{GPE}'. Similarly, we identify those entities using spaCy. These labels represent potential subjects of news articles. The insertion attack for NewsAPI aims to introduce three malicious entities, each corresponding to one of the targeted entity types. Substitution attacks replace these entities with the aforementioned malicious entities, while deletion attacks remove the entities entirely. \\More details are listed in Appendix~\ref{tab:modificationDetails}.


\subsection{Experimental Results}
\paragraph{WeatherAPI} We first evaluate the vulnerability of WeatherAPI. \tabref{tab:weatherapi} showcases the varying levels of susceptibility among models to different adversarial tactics. The bolded results indicate the highest ASR, while the underlined results represent the second highest ASR.
Generally, LLMs have showen greater vulnerability to substitution attacks than deletion attacks, indicating they struggle more with processing misleading or incorrect data than with the absence of information. 
Insertion attacks, which entail embedding irrelevant data into the API responses, were less effective across all models, as indicated by lower ASRs. This suggests that such attacks are more challenging to execute successfully. However, even a moderate level of success in these attacks has significant implications for the reliability of models in real-world applications. Additionally, Gemini was shown to be more vulnerable compared to GPT3.5-turbo in all three types of adversarial attacks we have conducted. 

\begin{table}[]
    \centering
    \scalebox{0.95}{
    \renewcommand{\arraystretch}{1.3}
    \begin{tabular}{c|c|c|c|cc|c}
        \toprule
        \multirow{2}{*}{\textbf{Model}} & \multirow{2}{*}{\textbf{Modified Field}} & \textbf{Deletion} & \textbf{Insertion} & \multicolumn{3}{c}{\textbf{Substitution}}\\
        \cline{3-7}
        & & \textbf{ASR}& \textbf{ASR}& \textbf{Deletion} & \textbf{Insertion}& \textbf{ASR}  \\
        \cline{1-7}
        \multirow{3}{*}{\textbf{GPT3.5-turbo}} & location & \textbf{93.10} & 57.24  & 89.65 & 91.72 & \underline{90.68}\\
        \cline{2-7}
        & temperature &\underline{86.67} & 60.33  & 93.33 & 90.33 & \textbf{91.81}\\
        \cline{2-7}
        & location + temperature & \underline{88.37} & 64.48 & 90.70 & 96.45 & \textbf{93.48} \\
        \cline{1-7} 
        \multirow{3}{*}{\textbf{Gemini}} & location & \textbf{91.30} & 73.53 & 86.95 & 93.52& \underline{90.12}  \\
        \cline{2-7}
        & temperature & \textbf{100.0} & 73.26  & 90.33 & 91.33 & \underline{92.32}\\
        \cline{2-7}
        & location + temperature & \textbf{90.32} & 79.08 & 90.32 & 89.10 & \underline{89.70} \\
        \bottomrule
    \end{tabular}}
    \caption{ASRs for WeatherAPI under third-party API attacks. We \textbf{bold} the highest ASR and \underline{underline} the second highest ASR for each row. 
    }
    \label{tab:weatherapi}
\end{table}

\begin{table}[]
    \centering
    \scalebox{1.0}{ 
    \renewcommand{\arraystretch}{1.3}
    \begin{tabular}{c|c|c|c|cc|c}
        \toprule
        \multirow{2}{*}{\textbf{APIs}} & \multirow{2}{*}{\textbf{Models}} & \textbf{Deletion} & \textbf{Insertion} & \multicolumn{3}{c}{\textbf{Substitution}}\\
        \cline{3-7}
        & & \textbf{ASR}& \textbf{ASR}& \textbf{Deletion} & \textbf{Insertion}& \textbf{ASR}  \\
        \cline{1-7}
        \multirow{2}{*}{\textbf{MediaWiki API}} & \textbf{GPT3.5-turbo} & \underline{70.60} & 30.10  & 87.80 & 63.30 & \textbf{73.56}\\
        \cline{2-7}
        & \textbf{Gemini} & \underline{55.79} & 33.72 & 74.19 & 55.91 & \textbf{63.77} \\
        \cline{1-7}
        \multirow{2}{*}{\textbf{NewsAPI}} & \textbf{GPT3.5-turbo} & \textbf{91.98} & 11.76 & 95.26 & 75.79 & \underline{84.42} \\
        \cline{2-7}
        & \textbf{Gemini} & \textbf{80.69} & \ \ 8.22 & 88.73 & 57.75 & \underline{69.96}  \\
        \bottomrule
    \end{tabular}}
    \caption{ASRs for MediaWiki API and NewsAPI under third-party API attacks. We \textbf{bold} the highest ASR and \underline{underline} the second highest ASR for each row. 
    }
    \label{tab:wiki_and_news_api}
    \vspace{-0.6cm}
\end{table}

\paragraph{MediaWiki and News APIs} We present the results of MediaWiki and News APIs in~\tabref{tab:wiki_and_news_api}. The insertion attack demonstrates significantly lower efficacy, especially with the News API. In contrast, the substitution and deletion attacks maintain high effectiveness, highlighting the LLMs' vulnerability to these attacks. Notably, the ASR difference between insertion and deletion underscores the greater challenge in embedding malicious content than removing it from LLM responses. The substitution attack, in particular, poses a greater threat to the MediaWiki API than the other two attacks. Further analysis reveals that within substitution attacks, deletion operations prove more effective than insertions, corroborating our findings on the performance of insertion and deletion attacks. For NewsAPIs, although the ASR for substitution attacks is lower than for deletion, it significantly exceeds that of insertion attacks.

\subsection{Discussion}

Based on the experiment results, we analyze and summerize several factors that affect the attack performance as follows. Table~\ref{tab:my_label} also provides some samples where attacks are unsuccessful among the majority of the successful attacks, and our explanations. 

\textbf{Conflicting Knowledge Injection}: If the manipulated information contradicts the LLM's internal knowledge, it might resist the attack, while LLMs might be more susceptible when they lack sufficient internal knowledge to identify inconsistencies. For example, removing location information makes it difficult for LLMs to identify the correct location, leading to a successful attack, but if the remaining weather data is unique to a specific region, the LLM might still be able to make an accurate guess.

\textbf{Reasoning Capabilities}: LLMs with better reasoning capabilities are more likely to resist attacks by identifying and disregarding inconsistencies, while LLMs with weaker capabilities are more susceptible to manipulation. For example, inserting a random temperature data point might be disregarded by the LLM, while subtly changing the temperature by a few degrees might be successful.

\begin{table}[]
    \centering
    \scalebox{0.9}{
    \renewcommand{\arraystretch}{1.3}
    \begin{tabular}{p{0.18\linewidth}|p{0.33\linewidth}|p{0.22\linewidth}|p{0.23\linewidth}}
    \toprule
      \textbf{Questions}   &  \textbf{Documents} & \textbf{Answers} & \textbf{Explanations}\\
\midrule
When do midlife crises happen? & A midlife crisis is a transition of identity and self-confidence that can occur in middle-aged individuals, typically\colorbox{mycolor}{ \st{40 to 60}} \textcolor{attack}{\textbf{1 to 60}} years old. & Midlife crisis happen in middle-aged individuals, typically between the ages of \colorbox{mycolor}{40 and 60}. & LLMs tend to resist manipulations that are extremely unreasonable. \\
\cline{1-4}
When did they show cartoons in movie theaters? &\textcolor{attack}{[The related information is not included in the provided API response]}& Movie theaters have been showing cartoons \colorbox{mycolor}{since their inception}.& Without related information in the provided API response, LLMs tend to generate meaningless answers. \\
\cline{1-4}
What can change the world, according to the activist? & \textcolor{attack}{[The information that is strongly related to the question’s answer are some abstract expressions.]} & \colorbox{mycolor}{Political commitment} and \colorbox{mycolor}{financial support} to achieve global development goals. & It's difficult for attackers to target key words on an abstract level that can influence the output of LLMs. \\

    \bottomrule
    \end{tabular}}
    \caption{Samples of questions, retrieved documents, answers, and explanations.}
\vspace{-0.1in}
\label{tab:my_label}
\end{table}

\textbf{Attack Quality} 
The techniques used to conduct the attack can affect the experiment result. In our experimental setting, attack performance can be affected by the named entity recognition techniques we use.  
Also, third-party APIs provide large amounts of information to LLMs, which at the same time can obstruct the attacker's ability to locate and conduct attacks precisely in a systematic manner. 

\section{Conclusion}

Our paper presents three attacks on third-party APIs integrated into the LLM ecosystems. This integration becomes more perilous as LLMs are increasingly equipped with APIs to better respond to user requests by accessing up-to-date information, performing complex calculations, and invoking external services through their APIs. It also opens up more possibilities for research in security within LLM ecosystems, extending beyond the isolated language models and APIs. Future work involves various attack methods, the design of defense mechanisms targeting third-party API attacks, and security concerns arising from multiple third-party API interactions.

\bibliography{iclr2024_conference}
\bibliographystyle{iclr2024_conference}
\newpage
\appendix

\section{Related Work}

\paragraph{Large Language Models}
Enabling machines to understand and communicate human languages has been a long-standing challenge. A machine is believed to be intelligent by researchers if it passes the Turing Test~\citep{french2000turing}, which is a deceptive test of determining the grasp of human intelligence by distinguishing whether a human or a machine generates the output. Language models (LMs) are one of the major techniques to model the generation of human languages based on the likelihood of words and sequences and make predictions of either the masked tokens in word sequences or future tokens. Large language models (LLM) are the most state-of-the-art LM that draw extensive attention due to their powerful capabilities. It scales the pre-trained LM such as BERT~\citep{devlin2018bert}, ELMo~\citep{peters2018deep} in terms of the training data size and model size. These transformer-based LMs offer a great improvement in advancing machine intelligence for various downstream tasks, which introduces transformative changes in both industry practices and daily use domains by providing high-quality responses to the user according to the given context. 

\paragraph{Third-Party API} 
Integrating third-party APIs with LLMs has been a pivotal advancement in AI, enabling these models to significantly extend their capabilities and applicability in the real world. This combination of recent works highlights a two-fold emphasis on broadening abilities and tackling emerging challenges.
Recent works such as Toolformer \citep{DBLP:journals/corr/abs-2302-04761}, ToolLLM \citep{qin2023toolllm}, API-Bank \citep{li-etal-2023-api}, and RestGPT \citep{song2023restgpt} demonstrate the potential of LLMs to autonomously leverage external tools and APIs, thus broadening their operational scope across various domains. 
LATM \citep{cai2023large}, which enables LLMs to create their own tools, and GeneGPT's \citep{jin2023genegpt} domain-specific applications illustrate the expanding problem-solving capacities and efficiencies of LLMs. 
The integration process also brings significant security and ethical considerations to the forefront. Efforts by Gorilla \citep{patil2023gorilla} and LLMDet \citep{wu-etal-2023-llmdet} to refine the precision of API calls and identify model-generated content underscore the critical need for mechanisms that ensure the responsible use of AI technologies. These contributions emphasize the importance of developing robust frameworks to mitigate risks associated with misinformation, misuse, and data privacy in deploying LLMs with third-party APIs.


\paragraph{Attacks} The surprisingly advanced performance of LLMs provides users with various high quality services, even security vulnerabilities detection for code repository \citep{yao2023survey}. However, the security concerns should be emphasised. \citet{li2020bert} proposed to use another LM BERT to mislead other deep neural models. The security risk of LLM is stressed by many prompt-based adversarial attacks. These attacks are mainly conducted by injecting pre-constructed prompts to LLMs in order to deceive LLM \citep{liu2023prompt, xu2023llm}. Apart from that, \citet{wang2023decodingtrust} attacks the performance of LLMs regarding the classical text classification task for GLUE dataset by leveraging an adversarial GLUE dataset AdvGLUE++. \citet{pedro2023prompt} evaluates the vulnerabilities of LLM-integrated web applications under the attacks which executes SQL injection through prompt injection. In addition, The security and privacy issues of OpenAI’s ChatGPT plugins are systematically assessed by \cite{iqbal2023llm}, with their constructed framework applied.


\section{Evaluation Metrics Description} \label{app:metrics}
We present the calculation of ASR for each attack below:
\begin{itemize}
    \item 
    \textbf{ASR for Insertion:} This metric quantifies the attack's ability to integrate additional, misleading information into the model's responses. It is calculated by tracking the instances where extraneous elements are successfully inserted into the model's output and fail to be recognized as such by the model. \\
    \[\text{ASR} = \frac{\text{\# of Successful Insertions}}{\text{\# of Valid Instances}}\]
    \item 
    \textbf{ASR for Deletion:} This metric measures the attack's success in removing crucial information from the input the model fails to recall or compensate for in its response. It is determined by the ratio of instances where essential information is effectively deleted, and the model does not identify or correct the omission. \\
    \[\text{ASR} = \frac{\text{\# of Successful Deletions}}{\text{\# of Valid Instances}}\]
    \item 
    \textbf{ASR for Substitution:}  ASR for substitution is conceived to offer a balanced evaluation of the attack's overall ability to manage both the addition and omission of information. It is defined as the harmonic mean of ASRs for insertion and deletion, similar to the F1-Score used in statistical analysis for measuring a test's accuracy. \\
    \[\text{ASR} = \frac{2*\text{InsertASR}*\text{DeleteASR}}{\text{InsertASR} + \text{DeleteASR}}\]
\end{itemize}

For all attacks, a higher ASR indicates a more effective attack, demonstrating the model's vulnerability to the corresponding attacks.

\clearpage
\section{Response format} \label{app:format}
Sample WeatherAPI Response
\begin{lstlisting}[style=json]
{
  "location": {
    "name": "London",
    "region": "City of London, Greater London",
    "country": "United Kingdom",
    "lat": 51.52,
    "lon": -0.11,
    "localtime": "2021-02-21 8:42"
  },
  "current": {
    "temp_c": 11,
    "temp_f": 51.8,
    "is_day": 1,
    "condition": {
      "text": "Partly cloudy",
    },
    "wind_mph": 3.8,,
    "pressure_in": 30.3,
    "precip_in": 0,
    "humidity": 82,
    "air_quality": {
      "co": 230.3,
      "no2": 13.5,
    }
  }
}
\end{lstlisting}

Sample MediaWikiAPI Response
\begin{lstlisting}[style=json]
{
    "batchcomplete": "",
    "query": {
        "pages": {
            "368118": {
                "pageid": 368118,
                "ns": 0,
                "title": "Madden NFL",
                "extract": "Madden NFL (known as John Madden Football until 1993) is an American football sports video game series developed by EA Tiburon for EA Sports. The franchise, named after Pro Football Hall of Fame coach and commentator John Madden, has sold more than 130 million copies as of 2018. Since 2004, it has been the only officially licensed National Football League (NFL) ..."
            }
        }
    }
}
\end{lstlisting}

\clearpage

Sample NewsAPI Response
\begin{lstlisting}[style=json]
{
    "storyId": "./cnn/stories/f382e1ca273b84cf5041d9ea589cd6d8c4651089.story",
    "text": "(CNN) -- A South Florida teenager accused of killing and mutilating 19 cats excitedly described to police how he dissected cats in class, and where to find cats for experimentation, according to police.\n\n\n\nTyler Weinman laughed when police told him they had information he was the cat killer, an arrest document said.\n\n\n\nWhen Miami-Dade police told Tyler Hayes Weinman someone was killing cats in the neighborhood..."
}
\end{lstlisting}

\section{Modification Rules}
\begin{table}[h]
\centering
\scalebox{0.7}{
\renewcommand{\arraystretch}{1.3}
\begin{tabular}{c|p{0.1\linewidth}||c|c|c}
\toprule
              & \textbf{Entity \newline  Label}       & \textbf{Insertion}                                                                                                             & \textbf{
              Substitution}                                                                        & \textbf{Deletion}                 \\ 
\cline{1-5}
Weather API   &  location & Introduce `not' before  & location: ``Sydney'' & Directly remove entities                         \\ 
\cline{1-5}
MediaWiki API & DATE               & Introduce 'not' before                                                                                                & \begin{tabular}[c]{@{}c@{}}Years: 1937, Dates: 1\end{tabular}                      & Directly remove entities \\ 
\cline{1-5}
NewsAPI       & ORG | GPR | PERSON & \begin{tabular}[c]{@{}c@{}}PERSON \& ORG: Insert `and Taishan' after\\ GPE: Insert `and Melbourne' after\end{tabular} & \begin{tabular}[c]{@{}c@{}}PERSON \& ORG: `Taishan'\\ GPR: `Melbourne'\end{tabular} & Directly remove entities \\ 
\bottomrule
\end{tabular}}
\caption{Entities and attack rules for the insertion, substitution, and deletion.}
\label{tab:modificationDetails}
\end{table}

\end{document}